\documentclass[prd,reprint,amsmath,nofootinbib,superscriptaddress,showpacs]{revtex4-1}
\usepackage[T1]{fontenc}
\usepackage[utf8]{inputenc}
\usepackage[brazil,english]{babel}
\usepackage{tensor}
\usepackage[usenames,dvipsnames]{color}
  \definecolor{darkblue}{RGB}{0,0,150}
\usepackage{amssymb} 
\usepackage{array}
\usepackage{amsthm} 
\usepackage{tensor}
\usepackage{graphicx}
\usepackage{nicefrac}
\usepackage{hyperref}
\usepackage{setspace}
\hypersetup{
  pdfinfo = {
    Author = {Alan M. da Silva, Daniel C. Guariento, C. Molina},
    Title = {Cosmological black holes and white holes with time-dependent mass}
  },
  colorlinks = true,
 linkcolor = darkblue,
 urlcolor = darkblue,
 citecolor = darkblue
}
\usepackage{multirow}
\usepackage{enumerate}
\usepackage{subcaption}

\newtheorem{theorem}{Theorem}[section]

\newtheorem{proposition}[theorem]{Proposition}
\newtheorem{corollary}[theorem]{Corollary}

\newcommand{\ud}{\ensuremath{\mathrm{d}}}

\DeclareMathAlphabet{\mathpzc}{T1}{pzc}{m}{it}

\def\coxa{{\Huge
C\kern-.1667em\lower.5ex\hbox{O}\-X\kern-.1667em\lower.5ex\hbox{A}\@}%
\index{CoXa}
}

\definecolor{darkorange}{rgb}{1.0,0.546,0.0}

\begin{document}

\title{Cosmological black holes and white holes with time-dependent mass}

\author{Alan M. da Silva}
\email{amsilva@if.usp.br}

\affiliation{Instituto de Física, Universidade de São Paulo,\\
Caixa Postal 66.318, 05315-970, São Paulo, Brazil}

\author{Daniel C. Guariento}
\email{carrasco@if.usp.br}

\affiliation{Instituto de Física, Universidade de São Paulo,\\
Caixa Postal 66.318, 05315-970, São Paulo, Brazil}

\author{C. Molina}
\email{cmolina@usp.br}

\affiliation{Escola de Artes, Ciências e Humanidades, Universidade de São Paulo,\\
Av. Arlindo Bettio 1000, 03828-000, São Paulo, Brazil}

\begin{abstract}

We consider the causal structure of generalized uncharged McVittie spacetimes with  increasing central mass $m (t)$ and positive Hubble factor $H (t)$. Under physically reasonable conditions, namely, a big bang singularity in the past, a positive cosmological constant and an upper limit to the central mass, we prove that the patch of the spacetime described by the cosmological time and areal radius coordinates is always geodesically incomplete, which implies the presence of event horizons in the spacetime. We also show that, depending on the asymptotic behavior of the $m$ and $H$ functions, the generalized McVittie spacetime can have a single black hole, a black-hole/white-hole pair or, differently from classic fixed-mass McVittie, a single white hole. A simple criterion is given to distinguish the different causal structures.

\end{abstract}


\maketitle

\section{Introduction}

Fully time-dependent solutions of general relativity may behave radically differently from their stationary counterparts. Familiar properties assumed for most solutions, many of which require exact time independence and asymptotic flatness at heart, do not hold on the more general dynamical scenarios \cite{Chadburn:2013mta,Faraoni:2007es,Abdalla:2006vb}. This breakdown of stationary properties on non-stationary spacetimes jeopardizes the interpretation and our very understanding of generic features of familiar spacetimes that we usually take for granted.

Moreover, the search for quantum gravity, as well as the resolution of the most pressing theoretical difficulties faced by cosmology, such as the cosmological constant problem and its many related issues, has spawned a rapidly expanding collection of modifications of general relativity, many of which can be interpreted in the Einstein frame as additional fields on top of the classical Einstein-Hilbert action. These theories have been found to yield several physically interesting non-vacuum solutions of the Einstein equations, many of which are by construction non-static. Even stationary solutions have been raising debates on such established results as the no-hair theorem \cite{Sotiriou:2013qea,*Sotiriou:2014pfa}, prompting for a revisit of such results and a renewed interest in checking its ranges of validity, as well as the precision of its statements.

In the study of the causal structure of non-vacuum solutions of general relativity, many concessions need to be made when stationarity is fully abandoned. Global properties have to be forfeit in favor of local ones, and inferences on the asymptotic structure need to take the bulk behavior into account.

One such example of a class of exact solutions to a modified gravity problem whose properties can be accessed via analytical means is the McVittie metric \cite{McVittie:1933zz}, which describes a black hole in an asymptotically FLRW universe. It was recently found \cite{Abdalla:2013ara} that it is a solution to the incompressible limit of $k$-essence minimally coupled to general relativity, also known as the \emph{cuscuton} field \cite{Afshordi:2006ad,*Afshordi:2007yx}. Some generalizations of the McVittie metric have been proposed in the literature, such as a charged central object \cite{shah-1968,*Faraoni:2014nba} and a time-dependent central mass \cite{Faraoni:2007es}. The latter has become known as the generalized McVittie metric (hereby referred to as gMcVittie for brevity) and has also been shown to be an exact solution of Einstein's equations for a scalar field coupled to gravity \cite{Afshordi:2014qaa}, specifically a particular case of the Horndeski theory \cite{Horndeski:1974wa} which can also be interpreted as an imperfect fluid in which a radial heat flow accounts for the increase in the central object's mass \cite{Guariento:2012ri}. It is related to the original fixed-mass McVittie metric via a disformal transformation, in which case the transformed action belongs to the more general $G^3$ class of scalar fields \cite{Gleyzes:2014qga,*Gleyzes:2014dya}.

The richness of the original McVittie solution's causal structure has been explored in previous works. It was shown that this metric actually describes a black hole in the appropriate limits \cite{Kaloper:2010ec,Lake:2011ni}, and also that the history of the free functions present in the metric is responsible for the asymptotic behavior \cite{daSilva:2012nh}. In the generalized case the situation becomes much more complex due to the addition of one more almost arbitrary time dependency coming from the mass function, and the analysis needs to be done without the simplifying properties of a constant central mass.

In this work we determine whether the physically interesting limits of the gMcVittie metric are geodesically incomplete and examine the number and nature of the apparent and event horizons of this spacetime. We find that when the expansion tends asymptotically to de~Sitter as time progresses, and when the central mass tends to a finite upper value, the gMcVittie spacetime is always geodesically incomplete. Moreover, whereas the original McVittie metric had two distinct asymptotic structures depending on the expansion history, namely a black hole and a black-hole/white-hole pair, we find that the generalized case presents a third possible outcome, in which only the white-hole horizon is present.

The paper is organized as follows: in Sec.~\ref{sec:structure} we review the apparent horizon structure of the gMcVittie metric and show that under physically meaningful assumptions these spacetimes are geodesically incomplete with respect to ingoing null radial trajectories, which implies that the geometries have event horizons that can be associated with either a black hole or a white hole; in Sec.~\ref{sectioncausal} we determine on which of these categories the event horizon falls, according to its dependence with the bulk history of the mass and expansion functions; we present our conclusions in Sec.~\ref{sec:conclusions}. Throughout the paper, derivatives with respect to the $t$ coordinate are denoted with an overhead dot. We use signature $(-,+,+,+)$ and natural units with $G=c=1$.

\section{Structural analysis of gMcVittie spacetimes} \label{sec:structure}

\subsection{General structure}

The uncharged gMcVittie metric, characterized by a mass function $m(t)$ and a scale factor $a(t)$, is given in areal radius and cosmological time coordinates by
\begin{equation}\label{metricaEF}
  \ud s^2 = - R^2 \ud t^2 + \left\{ \frac{\ud r}{R} - \left[ H + M \left( \frac{1}{R} - 1 \right) \right] r \ud t \right\}^2 + r^2 \ud \Omega^2 \,,
\end{equation}
%
%

with
\begin{align}
  R (t,r) \equiv&\, \sqrt{1 -\frac{2 m (t)}{r}} \,, \label{Rdef} \\
  H (t) \equiv&\, \frac{\dot{a}}{a} \,,\\
  M (t) \equiv&\, \frac{\dot{m}}{m} \, .
\end{align}
%
We impose the condition that $M < H$ at all times so that the causal structure of the singular surface $r = 2 m$, namely the fact that it is spacelike everywhere and to the past of all causal curves, is retained from the standard McVittie metric \cite{Guariento:2012ri}. The motivation for this somewhat restrictive condition is to keep the ``McVittie big bang'' a regular Cauchy surface from which to formulate a well behaved initial-value problem that allows us to use null geodesics to build the causal structure of the spacetime \cite{Lake:2011ni,walker-1970}.

Apparent horizons are defined as the surfaces in which congruences of null geodesics change their focusing properties\footnote{Here we are using the expression ``apparent horizon'' as a synonym of ``trapping horizon'' as coined in Ref.~\cite{Hayward:1993mw}. This coincides with the use of the term in Ref.~\cite{Senovilla:2011fk}, for example, but it is not equivalent to the definition given in Ref.~\cite{hawking}, since the latter is not a quasi-local definition.}. This corresponds to surfaces where
\begin{equation} 
  \Theta_{\text{in}} \Theta_{\text{out}} = 0 \,,
\end{equation}
where $\Theta_{\text{in/out}}$ are the expansions of spherical congruences along radial null directions (the subscript ``$\text{in}$'' refers to the ingoing direction while ``$\text{out}$'' refers to the outgoing direction). Regions where $\Theta_{\text{in}} \Theta_{\text{out}} < 0$ are called \emph{regular regions} and regions where $\Theta_{\text{in}} \Theta_{\text{out}} > 0$ are \emph{trapped regions} if $\Theta_{\text{in/out}} < 0$ or \emph{anti-trapped regions} if $\Theta_{\text{in/out}} > 0$.%
\footnote{This naming convention is found, for example, in Ref.~\cite{Dafermos:2004wr}. Some authors prefer to use ``normal'' or ``untrapped'' instead of ``regular'' and ``future trapped'' (``past trapped'') instead of ``trapped'' (``anti-trapped'').}
%

Setting $\ud s^2 = 0$ for null geodesics in Eq.~\eqref{metricaEF}, it immediately follows that the relevant equations are
%
%
%
\begin{equation}\label{apphor-r}
  \left( \frac{\ud r}{\ud t} \right)_{\text{in/out}} = R \left( r H \pm R \right) + r M \left( 1 - R \right) = 0 \,,
\end{equation}
where, due to our coordinate choice, $0 < R < 1$. Since the turning points of $\Theta_{\text{in/out}}$ coincide with the roots of $\left(\frac{\ud r}{\ud t} \right)_{\text{in/out}}$ \cite{Kaloper:2010ec,Faraoni:2012gz}, we can see here that for an accreting black hole ($M > 0$) in an expanding universe ($H > 0$), only the ingoing null geodesics, which correspond to the minus sign, may have real roots in a finite $(t,r)$ patch\footnote{This property is also observed in the Schwarzschild-de~Sitter metric if one switches from isotropic coordinates to areal radius coordinates. The outgoing horizons are brought to finite coordinates via a subsequent transformation in the $t$ coordinate, which the McVittie metric does not allow.}. Therefore in the analysis of the apparent horizon structure we only consider the minus sign.

If we use the definition of $R$ from Eq.~\eqref{Rdef} to substitute $r$ as the main variable for which to solve Eq.~\eqref{apphor-r}, we can cast it as
\begin{equation}\label{apphor-R-lr}
  R^4 - R^2 = - 2 m \left[ \left( H - M \right) R + M \right] \,.
\end{equation}
This is a fourth-order algebraic equation, whose real solutions in the interval $0 < R < 1$ ($r > 2 m$) correspond to the loci in which radial null geodesics have constant radius in the ingoing direction. For continuous $M$ and $H$, a spacelike hypersurface containing an odd-multiplicity solution corresponds to the existence of a boundary between an anti-trapped and a regular region, while even-multiplicity solutions separate regions which are either both anti-trapped or both regular. 
%

\begin{figure}[!htp]
  \centering
  \includegraphics[width=0.45\textwidth]{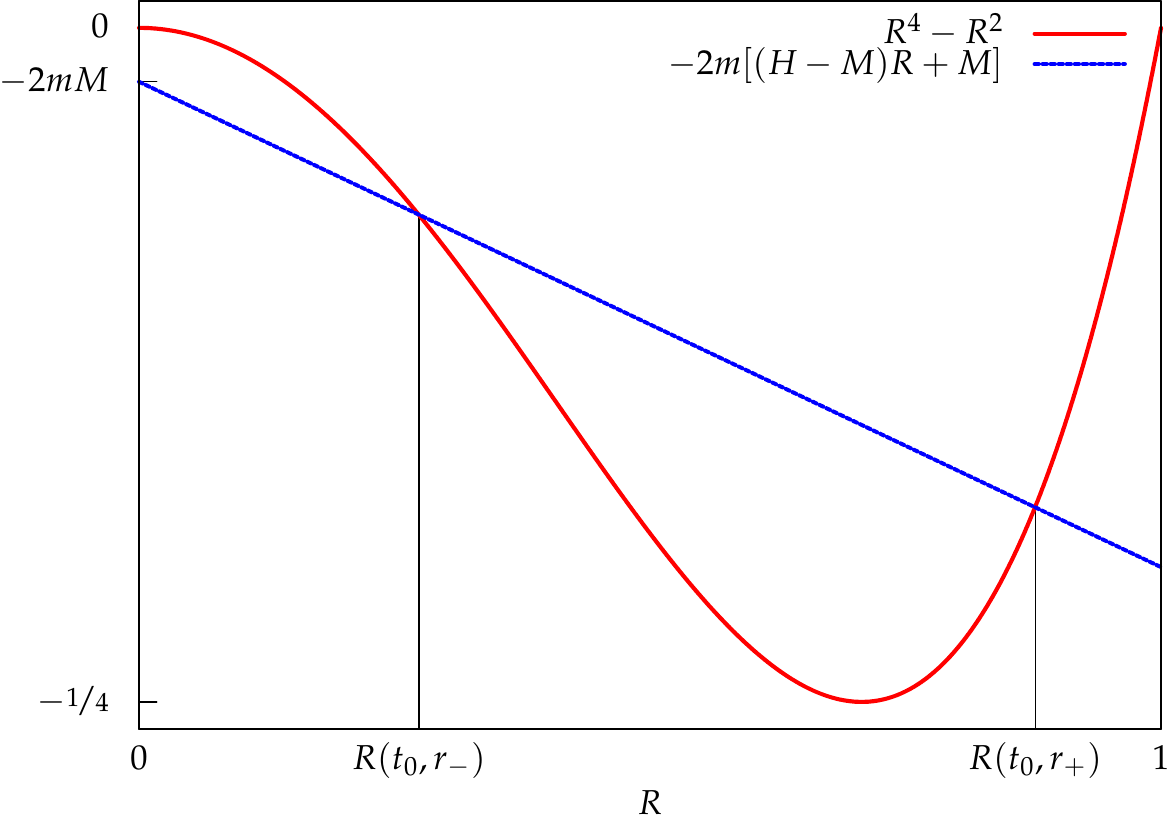}
  \caption{Visualization of the horizons in the physically relevant patch of generalized McVittie as intersections between the curves 
associated with the left- and right-hand sides of Eq.~\eqref{apphor-R-lr}.}
  \label{fig:apphor}
\end{figure}

Since there is no cubic term in Eq.~\eqref{apphor-R-lr}, all solutions must add up to zero, and there can be at most three positive real solutions \cite{stuchlik-1999}. Moreover, Eq.~\eqref{apphor-R-lr} is helpful for the determination of the number of real roots admissible in the range $0 < R < 1$, which is the physically interesting one. If we think of the horizons as the intersection between the curve represented in the left-hand side with the straight line in the right-hand side, it becomes manifest that there can be at most two positive roots in the allowed range for $R$, as in the McVittie metric. For that reason, we adopt the same naming convention and denote the inner apparent horizon as $r_- (t)$ and the outer horizon as $r_+ (t)$. It can also be seen that, for certain histories of $M$ and $H$, pairs of horizons may appear and disappear as the spacetime evolves, even if the restriction $M < H$ is always respected \cite{fontanini-ksm-2013}. Figure~\ref{fig:apphor} illustrates this structure for one particular 
example of $M$ and $H$ at a fixed time slice.

This horizon structure implies that, if we restrict our analysis to gMcVittie spacetimes that tend to non-extremal Schwarzschild-de Sitter as $t \to \infty$, we are guaranteed to have a regular region bounded by two apparent horizons, $r_-$ and $r_+$, for sufficiently large times. At large $r$, the geometry behaves asymptotically as a Friedmann-Lemaître-Robertson-Walker (FLRW) metric, which means that all future-directed outgoing null geodesics extent to null infinity, since they always reach distances sufficiently large to be well approximated by outgoing geodesics in FLRW space. On the other hand, future-directed ingoing null geodesics that start in the normal region tend to fall towards $r_-$, accumulating in its neighborhood as $t \to \infty$ at radius
\begin{equation}
r_\infty \equiv \lim_{t \to \infty} r_-(t) \, .
\end{equation}
It is this limit surface $(t \to \infty, r_\infty)$, mapped by ingoing null geodesics, that may constitute a black-hole horizon. If ingoing null geodesics departing from some initial event reach $t \to \infty$ in a finite affine parameter, it means that the patch of spacetime described by Eq.~\eqref{metricaEF} is geodesically incomplete. This issue has been thoroughly discussed in the literature regarding the standard McVittie spacetime \cite{Kaloper:2010ec,Lake:2011ni} and special cases of gMcVittie in which the time derivative of the mass function has support in a finite interval \cite{Guariento:2012ri}, and in these cases it was shown that under certain conditions the surface $(t \to \infty, r_\infty)$ constitutes a traversable limit surface.

\subsection{Geodesic incompleteness}\label{sec:timeflight}

%
To write the geodesic equations in more physically meaningful terms, we use as a starting point the comoving flow in isotropic coordinates $u_{\mu} \equiv -R \ud t$, which corresponds to the Hubble flow in the asymptotically FLRW region. The expansion scalar $\Theta_{(u)}$ of the timelike flow $u_{\mu}$ is  given by 
\begin{equation}
  \Theta_{(u)} = 3 \left[H + M \left( \frac{1}{R} - 1 \right)\right] \equiv 3 \mathcal{H} (r,t) \,,
\end{equation}
so that $\mathcal{H}$ is defined as a generalization of the Hubble factor in the asymptotic analysis.  We then rewrite the radial null trajectory in Eq.~\eqref{apphor-r} as
\begin{equation}\label{t of r}
  \frac{\ud r}{\ud t} = - R \left( R - \mathcal{H} r \right) \,.
\end{equation}
Using this result, the radial null geodesic equations may be cast as
\begin{align}
  \frac{\ud^2 r}{\ud \lambda^2} =&\, - \frac{\partial_t\left[R(R- \mathcal{H}r)\right]}{R^2 (R - \mathcal{H}r)^2 } \left(\frac{\ud r}{\ud \lambda} \right)^2 \, , \label{r of lambda} \\
  \frac{\ud^2 t}{\ud \lambda^2} =& - \left[ M + \frac{H - M}{2} \left( R + \frac{1}{R} \right) - \frac{2 m}{r^2} \right] \left( \frac{\ud t}{\ud \lambda} \right)^2 \,. \label{t of lambda}
\end{align}


Since the surface $(t \to \infty, r_\infty)$ is at a coordinate infinity, to assess the geodesic completeness of gMcVittie we need to check whether radial null geodesics starting from initial conditions at some event in the spacetime reach the surface $(t \to \infty, r_\infty)$ after a finite affine parameter. If this is the case, this surface is a causal horizon and the spacetime can be analytically extended beyond it. Otherwise, the surface is at an infinite affine distance from any point in the spacetime, so it is not an event horizon, but another patch of null infinity (along with the patch defined by outgoing null geodesics). The late-time behavior of ingoing null geodesics for large $r$ allows us to establish the following theorem:

\begin{theorem} \label{oteorema}

The patch of gMcVittie solutions described by metric \eqref{metricaEF} in the $(t, r)$ coordinates, with smooth $m(t)$ and $H(t)$  for all $t > 0$ and under the following hypotheses:
\begin{subequations}\label{hypotheses}
  \begin{align}
    m(t) > 0 \, ,\quad \forall t &> 0 \,, \label{mpositive} \displaybreak[0]\\
    m_0 \equiv \lim_{t \to \infty} m(t) &> 0\, ,\label{m0lim} \displaybreak[0]\\  
    H_0 \equiv \lim_{t \to \infty} H(t) &> 0 \,, \label{H0lim} \displaybreak[0]\\
    \frac{1}{3 \sqrt{3}} > m_0 H_0 &> 0 \,, \label{NonExtremalCondition} \displaybreak[0]\\
    M(t) > 0 \,, \, \forall \, t &> 0 \,, \label{AccretingCondition} \displaybreak[0]\\
    H(t) -  M(t) > 0 \,, \, \forall \, t &> 0 \,, \label{BigBangCondition}
  \end{align} 
\end{subequations}
is null-geodesically incomplete.

\end{theorem}

Each of the hypotheses grouped in Eqs.~\eqref{hypotheses} of the theorem has a physical motivation. Eq.~\eqref{H0lim} means that there is an expanding cosmological background; Eq.~\eqref{m0lim} means that the mass is bounded; Eq.~\eqref{NonExtremalCondition} is interpreted as the spacetime having a non-extremal Schwarzschild-de~Sitter limit as $t \to \infty$; Eq.~\eqref{AccretingCondition}  corresponds to an accreting central object; and Eq.~\eqref{BigBangCondition} means that the spacetime presents a spacelike singularity in the past. Under these assumptions, Theorem \ref{oteorema} guarantees that all gMcVittie spacetimes are null geodesically incomplete and have a traversable surface at $(t \to \infty, r_\infty)$. Note that Eq~\eqref{BigBangCondition} excludes the Sultana-Dyer spacetime \cite{Sultana:2005tp}, which corresponds to a gMcVittie space with $H = M$. As a particular case of this result, we include all standard McVittie spacetimes satisfying $H(t)> 0$, $H_0 > 0$ and $\frac{1}{3\sqrt{3}} > mH_0 > 0$, which generalizes\footnote{In \cite{Kaloper:2010ec} the additional assumption that $\dot{H}(t) < 0$ was made.} the results of \cite{Kaloper:2010ec}.

Since ingoing null geodesics are uniquely determined by any two of Eqs.~\eqref{t of r}, \eqref{r of lambda} or \eqref{t of lambda}, we choose to work with Eqs.~\eqref{t of r} and \eqref{r of lambda}. From Eq.~\eqref{t of r}, we note that ingoing null geodesics $r (t)$ always approach the inner horizon $r_-(t)$, since $R- \mathcal{H} r > 0$ above $r_-$ (unless we are also above $r_+$) and $R - \mathcal{H}r < 0$ below $r_-$. This means that all ingoing radial null geodesics that contain events in the regular region at sufficiently large times tend to $r_\infty$ as $t \to \infty$, as well as those that are below $r_-$ at sufficiently large times. Hypotheses \eqref{m0lim}, \eqref{H0lim}, \eqref{NonExtremalCondition} and \eqref{AccretingCondition} guarantee that there exists $T$ such that we have the two apparent horizons $r_-$ and $r_+$ for all $t > T$. With those observations in mind we are ready to present the proof of Theorem~\ref{oteorema}, by computing the leading term of the expression for $\Delta \lambda$ for an ingoing null geodesic taken near $r_\infty$ in the limit as $t \to \infty$.

\begin{proof}
In order to study the asymptotic behavior, we can take $t$ large enough so that we can linearize our expression near the  $t \to \infty$ limit value of the functions. For economy of notation, we denote as $\delta$ the order of magnitude of our linear approximations:
\begin{equation}
  \delta~\equiv~\max\left( \frac{r-r_\infty}{r_\infty} , \Delta H r_\infty, M r_\infty, \frac{m-m_0}{r_\infty} \right) \,.
\end{equation}
We also define $R_\infty \equiv R(m_0, r_\infty)$ and $\Delta H(t) \equiv H(t) - H_0$. Using this notation we can approximate $R [m(t), r]$  at large times, up to $o (\delta)$ terms%
, by
\begin{equation}
  \begin{split}
    R [m(t), r] =&\, R_\infty + \partial_m R_\infty (m-m_0) \\
    &+ \partial_r R_\infty (r- r_\infty) + o (\delta ) \\
    =&\, R_\infty - \frac{m-m_0}{r_\infty R_\infty} + \frac{m_0}{r_\infty^2R_\infty}(r-r_\infty) + o (\delta) \,,
  \end{split}
\end{equation}  
and, since $\lim_{t \to \infty} M(t) = 0$ because of Eq.~\eqref{m0lim}, we can also write
\begin{equation}
  \begin{split}
    r\mathcal{H}(t,r) =&\, r_\infty H_0 +r_\infty \left[ \Delta H(t) + \frac{M}{R_\infty}\left(1- R_\infty \right)\right] \\
    &+ (r-r_\infty) H_0 + o (\delta) \,.
  \end{split}
\end{equation}
Recalling that $R_\infty - r_\infty H_0 = 0$ and $\frac{m_0}{r_\infty} = \frac{1-R_\infty^2}{2}$, we obtain at first order in $\delta$
\begin{equation}
  \begin{split}
    R - \mathcal{H}r =&\, \left( \frac{1 - 3 R_\infty^2}{2 R_\infty} \right) \left( \frac{r-r_\infty}{r_\infty} \right) - \frac{m - m_0}{r_\infty R_\infty} \\
    & - \left[ \Delta H + \frac{M}{R_\infty} \left( 1 -R_\infty \right) \right]r_\infty + o (\delta)\,.
  \end{split}
\end{equation}

Near $r_\infty$, Eq.~\eqref{t of r} becomes at order $\delta$
\begin{equation}
 \frac{\ud r}{\ud t} 
 = - R_\infty \left[ \alpha \frac{r-r_\infty}{r_\infty} - \xi(t) \right] + o (\delta) \,, \label{linearapprox}
\end{equation}
with
\begin{align}
  \xi (t) \equiv&\, \frac{m-m_0}{r_\infty R_\infty} + r_\infty \left[ \Delta H + \frac{M}{R_\infty} ( 1 - R_\infty ) \right] \, , \label{epsilon-def}\\
  \alpha \equiv&\, \frac{1 - 3R_\infty^2}{2R_\infty}>0 \,.
\end{align}
In order to simplify our expressions, we also define $z(t) \equiv r(t) - r_\infty$ analogously to Ref.~\cite{daSilva:2012nh}. In terms of $z$,  Eq.~\eqref{linearapprox} reads
\begin{gather}
 \frac{\ud z}{\ud t} = - \alpha H_0z + R_\infty \xi(t) + o (\delta) \,,\label{linearapproxZ}
\end{gather}
whose solution, with initial condition $z(t_0) = z_0$, is
%
\begin{equation}\label{expH0t}
 z(t) = R_\infty e^{-\alpha H_0 t} \int_{t_0}^t e^{\alpha H_0 t'} \xi(t') \ud t' + Z_0 e^{- \alpha H_0 t } + o (\delta) \,,
\end{equation}
where $Z_0 \equiv z_0e^{\alpha H_0 t_0}$.
Now, working on Eq. \eqref{r of lambda} near $r = r_\infty$ or small $z$, we obtain at order $\delta$
\begin{equation}
\begin{split}
  - \frac{\partial_t \left[ R (R- \mathcal{H}r) \right]}{R^2 (R- \mathcal{H}r)^2 } = &\, - \frac{ \partial_t \left\{ R_\infty \left[ \alpha \frac{r- r_\infty}{r_\infty} - \xi(t) \right]  \right\}}{R_\infty^2 \left( \alpha \frac{r- r_\infty}{r_\infty} - \xi(t) \right)^2} + o (\delta) \\
  =&\, \frac{ \dot{\xi}(t)}{R_\infty \left[\frac{\alpha }{ r_\infty}z -\xi(t) \right]^2} + o (\delta) \,,
\end{split}
\end{equation}
where
\begin{equation}
 \dot{\xi}(t) =\frac{1- R_\infty^2 }{2 R_\infty} M(t) + r_\infty \left[ \dot{H}(t)  + \frac{1- R_\infty}{R_\infty} \dot{M}(t) \right] \, . \label{epsilon ponto}
\end{equation}
We obtain for $z(\lambda)$
\begin{equation}\label{equationf}
 \frac{ \ud^2 z}{\ud \lambda^2} = \frac{ \dot{\xi}(t)}{R_\infty \left[ \frac{\alpha}{ r_\infty}z - \xi(t) \right]^2} \left( \frac{\ud z}{\ud \lambda} \right)^2 + o (\delta) \,. 
\end{equation}
Integrating the leading term of Eq.~\eqref{equationf} once, we find
\begin{align}
 \frac{\ud z}{\ud \lambda} =&\, K \exp \left\{ \frac{R_\infty}{\alpha^2H_0^2} \int \frac{\dot{\xi} (t)}{\left[ z - \frac{r_\infty}{\alpha} \xi(t) \right]^2} \ud z \right\} \,, \label{first integral}
\end{align}
where $K$ is a constant and $t$ should be understood as a function of $z$. 

Here we need to analyze the leading term of $z(t)$, which depends on the behavior of $\xi(t)$. Three distinct regimes are possible:

\begin{enumerate}
\item \label{epsilonpequeno} If $\xi(t) = o (e^{-\alpha H_0 t})$, we can choose $t$ large enough such that the solution reduces to
\begin{equation}
  z(t) =  A e^{- \alpha H_0 t} + o (e^{-\alpha H_0 t}) \, .\label{ZdoKaloper} \, 
\end{equation}
where the constant $A$ codifies the dependence of $z(t)$ on the initial conditions.
\item \label{epsilongrande} If $\xi(t) > \mathcal{O}(e^{-\alpha H_0 t})$, for large $t$, we have\footnote{More detail about this approximation can be found in appendix~\ref{appendix:flighttime}.} 
\begin{equation}
  z(t) = \frac{ r_\infty}{\alpha} \left[ \xi(t) - \frac{1}{H_0 \alpha}\dot{\xi}(t) \right] + o [\dot{\xi}(t)] \, . \label{Znosso} 
\end{equation}
\item \label{epsilonigual} If $\xi(t) = C_\infty e^{-\alpha H_0 t} + o (e^{-\alpha H_0 t})  $, where $C_\infty$ is a constant, we can solve the integral in Eq.~\eqref{expH0t} to find
\begin{equation}
 z(t) = R_\infty C_\infty t e^{-\alpha H_0 t} + \mathcal{O}(e^{-\alpha H_0 t})  \,. \label{outroZ}
\end{equation}
\end{enumerate}

We now work out each of these cases individually.

\begin{description}

\item[\textbf{Case~\ref{epsilonpequeno}}] We may use Eq.~\eqref{ZdoKaloper} to change the integration variable to $t$, which implies that $\ud z = -\left[ {A {\alpha}H_0} e^{-\alpha H_0 t} + o (e^{-\alpha H_0 t}) \right] \ud t$. Eq.~\eqref{first integral} then reads at leading order
\begin{equation}
  \frac{\ud z}{\ud \lambda} = K \exp \left[ - \frac{ R_\infty  }{\alpha H_0 A  } \int e^{\alpha H_0 t} \dot{\xi}(t) \ud t \right] \,.
\end{equation}
Solving for $\lambda$, we find
\begin{equation}
 K \Delta \lambda = -\int_{z_0}^{0}  \exp \left[ \frac{R_\infty}{\alpha H_0 A} \int e^{\alpha H_0 t} \dot{\xi}(t) \ud t \right] \ud z  \,, \label{deltalambda}
\end{equation}
and changing the integration variable to $t$ once more, we obtain at leading order
\begin{equation}
  K \Delta \lambda =
  A H_0 \alpha \int_{t_0}^{\infty} \exp \left[ \frac{R_\infty}{\alpha H_0 A } \int^t e^{\alpha H_0 t'} \dot{\xi}(t') \ud t' - \alpha H_0 t \right] \ud t  \, .
\end{equation}
Therefore, geodesic incompleteness is equivalent to the convergence of the expression
\begin{equation}
 I = \int_{t_0}^{\infty}  \exp \left[ \frac{r_\infty}{{\alpha} A} \int^t e^{\alpha H_0 t'} \dot{\xi}(t') \ud t' - \alpha H_0 t \right] \ud t \,, \label{criterion}
\end{equation}
where we have used the fact that $R_\infty = r_\infty H_0$ to further simplify the expression. Recalling that, in this case, $\xi(t) = o (e^{-\alpha H_0 t})$, then necessarily $\dot{\xi}(t) = o (e^{-\alpha H_0 t})$. This implies that $\left|\int^t e^{\alpha H_0 t'} \dot{\xi}(t') \ud t' \right| < \int^t e^{\alpha H_0 t'} e^{- \alpha H_0 t'} \ud t' = \mathcal{O}(t)$, so the leading term in the argument of the exponential in Eq.~\eqref{criterion} is the linear one. Thus, Eq.~\eqref{criterion} reads
\begin{equation}
 I = \int_{t_0}^\infty e^{- \alpha H_0 t'} \ud t' + o (e^{ - \alpha H_0 t}) \, ,
\end{equation}
which is convergent. Therefore, the ingoing geodesics reach the limit surface in a finite parameter time.


\item[\textbf{Case~\ref{epsilongrande}}] Here, $\xi(t)$ dominates the late-time behavior of the flow. The deduction works in the same way as for case~\ref{epsilonpequeno} up to Eq.~\eqref{first integral}, but when we express the integrand as a function of $t$ the expression behaves differently. Taking $z = \frac{ r_\infty}{\alpha} \xi(t) - \frac{R_\infty }{\alpha^2 H_0^2}\dot{\xi} + o (\dot{\xi}) $ and $\ud z = \frac{ r_\infty}{\alpha} \dot{\xi}(t) \ud t + o (\dot{\xi}) $, we find for the leading term
\begin{equation}
\begin{split}
  \frac{\ud z}{\ud \lambda} =&\, -K \exp\left\{ \frac{R_\infty }{ \alpha^2 H_0^2 } \int \frac{\dot{\xi}(t)}{\left[ z - \frac{ r_\infty}{\alpha} \xi(t) \right]^2} \ud z  \right\}  \\
  =&\, -K' e^{\alpha H_0 t}  \,.
\end{split}
\end{equation}
We then integrate again to obtain $\Delta \lambda$:
\begin{equation}
  \begin{split}
    K' \Delta \lambda =&\, \int_{z_0}^0 e^{-\alpha H_0 t } \ud z \\
    =&\, \frac{ r_\infty}{\alpha} \int_{t_0}^{\infty} e^{-\alpha H_0 t } \dot{\xi}(t) \ud t \, , 
  \end{split}
\end{equation}
so in this case geodesic incompleteness is equivalent to the convergence of the expression
\begin{equation}
 I' = \int_{t_0}^{\infty} e^{-\alpha H_0 t } \dot{\xi}(t) \ud t \, . \label{criterion2}
\end{equation}
However, the integral \eqref{criterion2} always converges when $\dot{\xi} \to 0$ as $t \to \infty$, which we are assuming.


\item[\textbf{Case~\ref{epsilonigual}}] With $\xi = C_\infty e^{-\alpha H_0 t} + o (e^{-\alpha H_0 t}) $, we have $z(t)$ given by Eq.~\eqref{outroZ}
, so that \[\ud z = - \left[R_\infty C_\infty\alpha H_0 t e^{-\alpha H_0 t}+ \mathcal{O}(e^{-\alpha H_0 t})\right] \ud t.\] Thus, Eq.~\eqref{first integral} takes the form
\begin{align}
  \begin{split}
    \frac{\ud z}{\ud \lambda} =&\, K \exp{\left\{ \frac{R_\infty }{ \alpha^2 H_0^2 } \int \frac{\dot{\xi}(t)}{\left[ z - \frac{r_\infty}{\alpha} \xi (t) \right]^2} \ud z \right\}} \\
    =&\, K' \,,
  \end{split}\\
  K'\Delta \lambda =&\, -z_0 \,.
\end{align}
\end{description}

This finishes our proof that in all cases satisfying hypotheses \eqref{hypotheses}, the surface $(t \to \infty, r_\infty)$ is always reachable by ingoing null radial geodesics in a finite affine parameter.
\end{proof}



\section{Causal Structure}\label{sectioncausal}

\begin{figure}[!htp]
  \centering
  \includegraphics[width=0.45\textwidth]{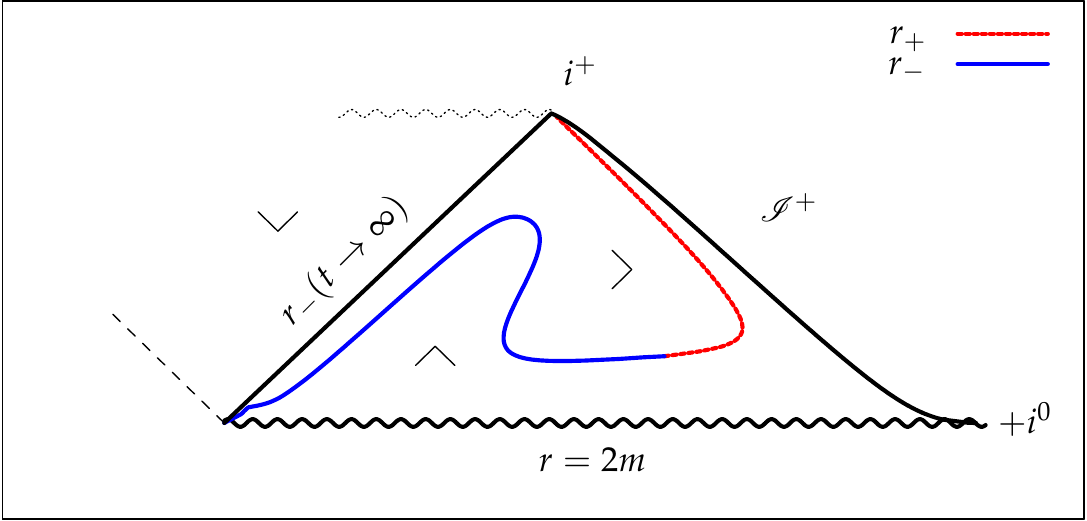}
  \caption{gMcVittie spacetime where all ingoing null geodesics (in blue) reach the limit surface from above the horizon.}\label{figurakaloper}
\end{figure}

\begin{figure}[!htp]
  \centering
  \begin{subfigure}{.45\textwidth}
    \includegraphics[width=\textwidth]{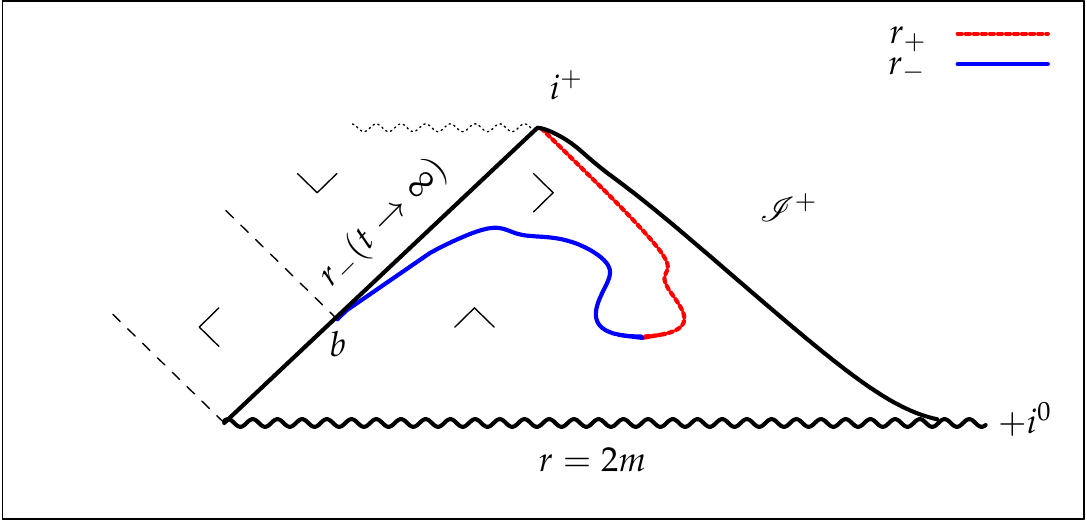}
    \caption{$\dot{\xi}(t) \to 0^{-}$} \label{figuralake}
  \end{subfigure}
  \begin{subfigure}{.45\textwidth}
    \includegraphics[width=\textwidth]{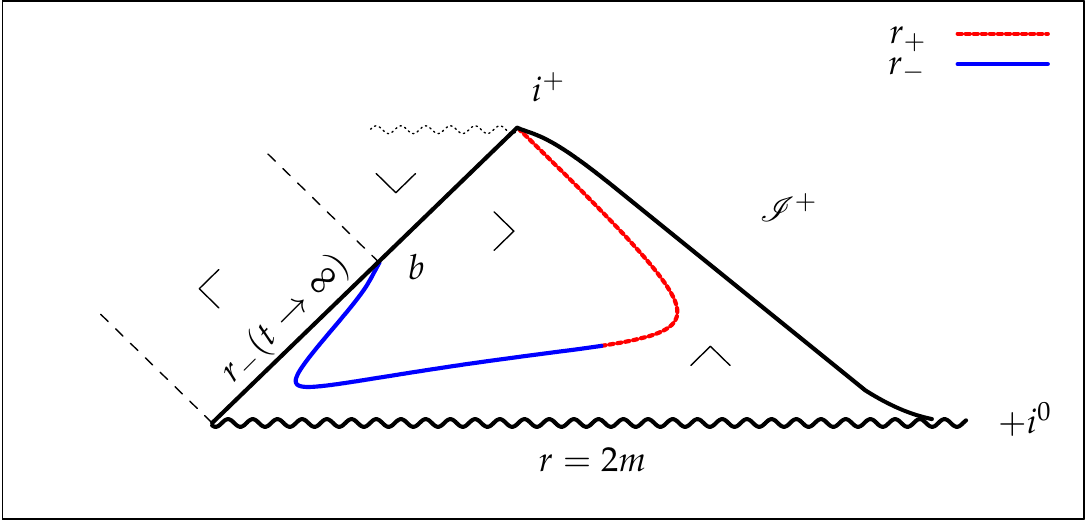}
    \caption{$\dot{\xi}(t) \to 0^{+}$}\label{figuralake-rpos}
  \end{subfigure}
\caption{gMcVittie spacetime where some ingoing null geodesics reach the $r_\infty$ limit  below $r_-$ (blue). Since later ingoing null curves emerge from the $r = 2m$ singularity at the left, we see that there is a limit time that separates curves that reach $r_\infty$ from above from curves that reach there from below.}
\end{figure}

\begin{figure}[!htp]
  \centering
  \includegraphics[width=0.45\textwidth]{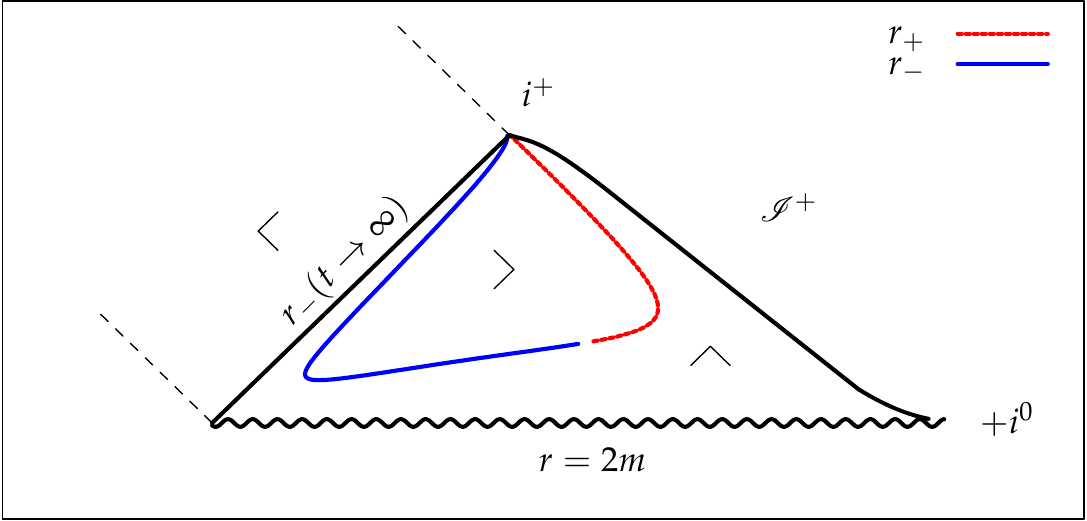}
  \caption{gMcVittie spacetime where all ingoing null geodesics reach the limit surface from below the $r_-$ apparent horizon (blue).} \label{figuragota}
\end{figure}

The results of Sec.~\ref{sec:structure} state that accreting gMcVittie spacetimes which satisfy the hypotheses \eqref{hypotheses} can be extended beyond the $(t \to \infty, r_\infty)$ limit surface. The natural question which arises in this situation is what kind of surface it is and what kind of region is hidden beyond it. In the standard McVittie scenario there has been a series of articles discussing this issue \cite{Kaloper:2010ec,Lake:2011ni,daSilva:2012nh}, showing that in McVittie the $(t \to \infty, r_\infty)$ limit surface can either be a black-hole event horizon or be composed of a black-hole horizon and a white-hole horizon. The distinction between these cases, as was shown in Ref.~\cite{daSilva:2012nh}, is how fast the ingoing null geodesics approach $r_\infty$ in relation to the apparent horizon $r_-$. If all geodesics approach the $r_\infty$ limit from above $r_-$, then the limit surface is a black-hole event horizon. If, on the other hand, there are ingoing null geodesics that approach the limit surface from below $r_-(t)$ at the same time as ingoing null geodesics that approach the limit surface from above $r_-$, then the limit surface contains a black-hole event horizon portion ``above'' $r_-$ as well as a white-hole event horizon portion, ``below'' $r_-$. These particular cases are depicted in Figs.~\ref{figurakaloper}, \ref{figuralake} and \ref{figuralake-rpos}, respectively\footnote{The model used to generate Figs.~\ref{figurakaloper}, \ref{figuralake}, \ref{figuralake-rpos} and \ref{figuragota} is $H (t) = H_0 \tanh^{-1} \left( \frac{3}{2} H_0 t \right)$ with $H_0 = 10^{-2}$, and $m (t) = m_0 + A \tanh \left( B H_0 t - 2.5 \right)$. Fig.~\ref{figurakaloper} uses $m_0 = 5$, $A = 0.5$ and $B = 5$; Fig.~\ref{figuralake} uses $m_0 = 8$, $A = 2$, $B = 5$; Fig.~\ref{figuralake-rpos} uses $m_0 = 8$, $A = 0.025$ and $B = 1.5$; and Fig.~\ref{figuragota} uses $m_0 = 10$, $A = 1$ and $B = 0.5$.}.

This distinction comes from the fact that the $r_\infty$ surface in the Schwarzschild-de Sitter spacetime, which is the limit of gMcVittie as $t \to \infty$ given \eqref{hypotheses}, is a zero of the expansion of outgoing null geodesics, $\Theta_{\text{out}}$. Since $\Theta_{\text{out}} > 0$ in the patch described by the $(t, r)$ coordinates, this means that behind the limit surface we have $\Theta_{\text{out}} < 0$. Then, the nature of the region behind the surface will depend on the sign of $\Theta_{\text{in}}$. Since between $r_-$ and $r_+$ we have $\Theta_{\text{in}} < 0$ (it is a regular region), then if it is possible for timelike or null observers to travel to the limit surface while being always above $r_-$ and reach it, then by continuity they will fall in a trapped region, as $\Theta_{\text{out}} \Theta_{\text{in}} > 0$ there. This is equivalent to saying that they have entered a black hole whose event horizon is the limit surface (or part of it). Conversely, below $r_-$  we have an anti-trapped 
region with $\Theta_{\text{out}} > 0$ and $\Theta_{\text{in}} > 0$. If we can cross the limit surface while always traveling below $r_-$, then by continuity we reach a  regular region where $\Theta_{\text{out}} < 0$ and $\Theta_{\text{in}} > 0$. This is equivalent to saying that we came out of a white-hole horizon. For more details in this argument, we refer the reader to Ref.~\cite{Lake:2011ni}.

In the following development, we will establish sufficient conditions over the functions defining each gMcVittie model, $m(t)$ and $H(t)$ that determine to which causal structures it corresponds. In this wider class of spacetimes, there is a new case, depicted in Fig.~\ref{figuragota}, in which all ingoing null curves approach the limit surface from below the apparent horizon $r_-$. In this case,  the limit surface is entirely a white-hole horizon. We can see in Figs.~\ref{figurakaloper}, \ref{figuralake}, \ref{figuralake-rpos} and \ref{figuragota} that each case corresponds to a different end point of the $r_-$ apparent horizon, when it reaches the $r_\infty$ limiting surface in the conformal diagrams. The causal structure is determined by the asymptotic behavior of the function $\xi(t)$, defined by Eq. \eqref{epsilon-def} according to Theorem~\ref{theorem-rpos}.

\begin{theorem}\label{theorem-rpos}

Let be a gMcVittie spacetime described by the metric \eqref{metricaEF} under the hypotheses \eqref{hypotheses} and, in addition, assume that the function $\dot{\xi}(t)$ defined in Eq.~\eqref{epsilon-def} tends to zero from positive (negative) values. Then, if there exists $\sigma > 0$ such that 
\begin{equation}
  F^{\sigma}_{-}(t_0, t) = \int_{t_0}^t e^{\left(\alpha H_0+ \sigma\right)u} \xi(u) \ud u \,, \label{Fmenos}
\end{equation}
converges as $t \to \infty$, then the gMcVittie spacetime presents a black hole and a white hole horizon. Analogously, if there exists $\bar{\sigma} > 0$ such that
\begin{equation}
 F^{\bar{\sigma}}_{+}(t_0, t) \equiv \int_{t_0}^t e^{\left(\alpha H_0 - \bar{\sigma}\right)u} \xi(u) \ud u \, \label{Fmais}
\end{equation}
diverges as $t \to \infty$, then the gMcVittie spacetime in question presents only a white hole (black hole) horizon.

\end{theorem}


\begin{table}[!htp]
  \caption{Possible asymptotic structures of gMcVittie spacetimes: summary of the results of Sec.~\ref{sectioncausal}.
} \label{tabela}
  \centering
\begin{tabular*}{\columnwidth}{*{3}{c@{\extracolsep{\fill}}}}
\hline \\
    & $ \dot{\xi}(t) \to 0^-$ & $\dot{\xi}(t) \to 0^+ $ \\ \\
    \hline \\
    \parbox[c]{0.5\columnwidth}{$ \left| \int^{\infty} e^{(\alpha H_0 - \sigma)u} \xi(u) \ud u \right| < \infty$} & \parbox[c]{0.2\columnwidth}{black hole\\ and \\ white hole} & \parbox[c]{0.2\columnwidth}{black hole \\ and \\ white hole} \\ \\
    \hline \\
    \parbox[c]{0.5\columnwidth}{$ \left| \int^{\infty} e^{(\alpha H_0 +  \sigma)u} \xi(u) \ud u \right| \to \infty$} & \parbox[c]{0.2\columnwidth}{black hole \\ only} & \parbox[c]{0.2\columnwidth}{white hole \\ only} \\ \\
\hline
  \end{tabular*}
\end{table}


The results can be more easily visualized in Table~\ref{tabela}. From Theorem~\ref{theorem-rpos}, one can see that the cases where gMcVittie contains only a white hole are restricted to cases where $\dot{\xi} \to 0^+$, that is, when it goes to zero from positive values. Symmetrically, the only black hole case correspond to $\dot{\xi} \to 0^-$. This is due to the fact that the sign of $\dot{\xi}$ is the sign of the slope of $r_-(t)$ for large $t$,
\begin{equation}
 \dot{r}_-(t) \approx \left. \frac{ \dot{\xi}(t)}{\partial_r (R- \mathcal{H}r)} \right|_{r= r_-(t)}\, , \label{dotrminus}
\end{equation}
since the denominator is positive at the $r_-$ horizon. 
We are assuming that the denominator is not degenerate, that is, $r_+$ and $r_-$ do not coincide. We can understand intuitively that when its slope is negative, it is easier for null rays to reach the horizon from above the apparent horizon, which correspond to the regular region, which leads to a black hole. In the same manner, when the slope of $r_-$ is positive, it is easier for null rays to traverse from below the apparent horizon, laying in the anti-trapped region, characterizing the limit of a white hole region. In both cases, if the absolute value of function $\xi(t)$ decreases fast enough, we have the case in which the limit surface correspond to a pair of white-hole/black-hole horizons, separated by a bifurcating two-sphere. The cases in which the limit surface has only one character are those in which the $\xi$ function does not decrease faster than the exponential that modulates it in Eq.~\eqref{Fmais}.

With these remarks in mind, we can build models corresponding to the four cases shown in Table~\ref{tabela}. First, we need to chose if the slope of $r_-$ will be positive or negative for large times, which by Eq.~\eqref{dotrminus} means choosing the sign of $\dot{\xi}$ for large times. Inspecting Eq.~\eqref{epsilon ponto}, we can see that the term proportional to $M$ is positive, by our initial assumptions. The sign of $\dot{H}$ can be either plus or minus, but physically realistic models usually correspond to $\dot{H} < 0$. This means that we can tune the functions $m(t)$ and $H(t)$ in such a way that the leading term for large $t$ is positive or negative. 


The second step is to choose how the leading term approaches zero. If we choose $m - m_0$ or $\Delta H$ (depending on what was chosen to be leading before) that approaches zero sufficiently faster than $\exp(-\alpha H_0 t)$, then there will surely exist $\sigma$ such that $F^\sigma_-$ converges and we are in the case where a pair black hole/white hole is present. If they decrease sufficiently slower than $\exp(-\alpha H_0 t)$, then we fall in the case were $F^\sigma_+$ diverges and there is only a black hole or white hole, depending on the former choice of sign for $\dot{\xi}$ being negative or positive, respectively. Recall that $\alpha$ and $H_0$ depend only on the limit values of the gMcVittie functions, but the causal structure depends on \emph{how} those limiting values are reached. Using the reasoning above we have built the models depicted in Figs.~\ref{figurakaloper}, \ref{figuralake}, \ref{figuralake-rpos} and \ref{figuragota}. 

The proof of Theorem~\ref{theorem-rpos} is analogous to the proof of the main result in Ref.~\cite{daSilva:2012nh} and the case-specific details can be found in Appendix~\ref{appendix:causal}. Roughly speaking, the proofs for both cases are based in the fact that we can substitute $r_-(t)$ by $ r = r_\infty$, which is a constant curve. Thus, we can evaluate the flow of ingoing null geodesics near $r = r_\infty$ and determine if ingoing null curves cross or fail to cross it for sufficiently large $t$. This is done by defining two families of approximative curves, one that approximates the ingoing geodesics from above and another that does so from below. The existence or non-existence of approximating curves that cross the $r = r_\infty$ line for arbitrarily large times leads to the criteria shown in Table~\ref{tabela}.
%


\section{Conclusions}\label{sec:conclusions}

In this work we have shown that all generalized McVittie spacetimes with sufficiently slow accretion, an expanding cosmological background, a big bang singularity in the past and a non-extremal Schwarzschild-de Sitter spacetime as its limit as $t \to \infty$ are geodesically incomplete, having a traversable surface located at $(t \to \infty , r_\infty)$. 

This result was proved by studying the leading term of the variation of the affine parameter of ingoing null geodesics as they approach the surface which accumulates near the inner apparent horizon $r_-(t)$. This was done by approximating the flow of the non-linear differential equation that governs them for large times, taking care to keep the right leading terms at every step. We have shown that, under the aforementioned assumptions, the ingoing geodesics always reach the surface in a finite affine parameter, implying that the patch of generalized McVittie spacetime, described by the cosmological time $t$ and the areal radius $r$ in Eq.~\eqref{metricaEF}, is null-geodesically incomplete. Our proof also includes standard McVittie spacetimes as a particular case with constant $m(t)$, since the asymptotic behavior of geodesics is governed by the asymptotic properties of $m(t)$ and $H(t)$ through the function $\xi(t)$, defined in Eq.~\eqref{epsilon-def}, and its time derivatives.

Therefore, in generalized McVittie spacetimes, the limit surface can be either a black hole event horizon, in part a black hole horizon and a white hole horizon, or even entirely a white hole horizon, depending again on the asymptotic properties of $\xi(t)$. We obtained this result by a similar method to the one used in \cite{daSilva:2012nh}, where the standard McVittie spacetime, with stricter assumptions than those used in this paper, was proved to allow two of the three possible causal structures found here, since the white-hole-only case is not possible in standard McVittie spacetimes that obey our assumptions. Following the observation that the causal structures are distinguished by the way ingoing null curves approach their limit at $r = r_\infty$ --- either all from above $r_-(t)$, all from below $r_-(t)$ or some from above and some from below --- our proof relies on the fact that we only need to study the crossing of the geodesics with respect to a constant $r=r_\infty$ surface, instead of the time-dependent $r_-(t)
$ horizon. We then built two families of approximations for the ingoing null curves: one family that approximates from above and the other one from below such that we could write formally the solution to those approximations in order to determine whether there exists one curve that crosses the $r = r_\infty$ surface, or, alternatively, whether all of them fail to cross it. This led us to the condition on $\xi(t)$ that determines the causal structure of almost all gMcVittie solutions (up to a set of zero measure) which satisfy the initial assumptions.

We are provided with a large family of analytical solutions that are dynamical and can present evolving black holes and white holes isolated or in pairs. Those can be useful for many applications, as a laboratory to the study of the physics of dynamical black holes, toy models for accreting astrophysical black holes, the study of bounded systems in cosmological backgrounds and, due to the richness of types of dynamical and causal horizons it presents, to the study of physics near different types of horizons, which are the scenario of many recent developments such as the fluid-gravity correspondence in dynamical backgrounds  \cite{Bhattacharyya:2008jc}, properties of vacuum solutions of modified gravity \cite{Berglund:2012fk,Gleyzes:2014dya,Gleyzes:2014qga}, the search for the thermodynamical laws of gravitational systems \cite{Hayward:1997jp, Acquaviva:2014owa} and even the debate on the proper definition of a black hole \cite{Hawking:2014tga,Frolov:2014wja,Visser:2014zqa}.

\begin{acknowledgments}
  We thank M. Fontanini for his contribution in the early stages of this work. A.\ M.\ S.\ is supported by FAPESP Grant No.\ 2013/06126-6. D.\ C.\ G.\ is supported by FAPESP Grant No.\ 2010/08267-8. C.\ M.\ is supported by FAPESP Grant No.\ 2012/15775-5 and CNPq Grant No.\ 303431/2012-1.
\end{acknowledgments}

\appendix
 \section{On the approximation used for $\xi(t) > \mathcal{O}(e^{-\alpha H_0 t})$} \label{appendix:flighttime}
 
 In section~\ref{sec:timeflight}, we claimed that the form of $z(t)$ when $\xi(t) > \mathcal{O}(e^{-\alpha H_0 t})$ was given by \eqref{Znosso}, which we used to prove geodesic incompleteness in case~\ref{epsilongrande}. Here we show the deduction of this result.
 
 Integrating the first term of Eq.~\eqref{expH0t} by parts, we obtain

\begin{equation}
  \begin{split}
    e^{-\alpha H_0 t} \int_{t_0}^t e^{\alpha H_0 t'} \xi(t') \ud t' =&\, \frac{1}{H_0 \alpha} \left[ \xi(t) - \xi(t_0)e^{- \alpha H_0 (t -t_0)} \right] \\
    &- \frac{ e^{- \alpha H_0 t}}{\alpha H_0} \int_{t_0}^{t} e^{\alpha H_0 t} \dot{\xi}(t')\ud t' \,,
  \end{split}
\end{equation}
which we may write iteratively as
\begin{equation}\label{epsilonserie}
  \begin{split}
    e^{-\alpha{ H_0 t}} \int_{t_0}^t e^{\alpha H_0 t'} \xi(t') \ud t' =&\, \sum_{n=0}^{\infty} (-1)^n\left( \frac{1}{H_0 \alpha} \right)^{n+1} \frac{\ud^n }{\ud t^n} \xi(t) \\
    &- C_{0} e^{-\alpha H_0 t} \,,
  \end{split}
\end{equation}
where $\frac{\ud^0 f}{\ud t^0} = f$ and
\begin{equation}
 C_0 = e^{\alpha H_0 t_0}\sum_{n=0}^{\infty} (-1)^{n}\left(\frac{1}{H_0 \alpha}\right)^{n+1} \frac{\ud^n }{\ud t^n}\xi(t_0) \,.
\end{equation}
Thus, we can write the flow of Eq.~\eqref{linearapproxZ} as
\begin{equation}\label{flow of z}
  \begin{split}
    z(t) =&\, (Z_0 - R_\infty C_0) e^{- \alpha H_0 t} \\
    &+ R_\infty \sum_{n=0}^{\infty} (-1)^n\left( \frac{1}{H_0 \alpha} \right)^{n+1} \frac{\ud^n }{\ud t^n} \xi(t) \\
    &+ o (\delta)\,.
  \end{split}
\end{equation}
Since Eq.~\eqref{epsilonserie} is meaninful only when the series converges, we use this result in cases where the $\xi$ dominates $\dot{\xi}$ and, consequently, $\dot{\xi}$ dominates higher derivatives of $\xi$. Thus, the leading term at late times is always the one with $n = 0$, and the first subleading term is $n=1$.  These assumptions corresponds to cases in which $\xi(t) = \textit{o}(e^{-\alpha H_0 t})$ (see Appendix \ref{sec:appendix}). The exceptions to this rule decrease either exponentially or faster and were treated separately in cases~\ref{epsilonigual} and \ref{epsilonpequeno}, respectively. 

\section{When $|\dot{\xi}(t)| > |\xi(t)|$ as $t \to \infty$}\label{sec:appendix}

In this Appendix we show why all the cases where approximation \eqref{Znosso} is valid are of slow decreasing, that is, correspond to case~\ref{epsilongrande} in the analysis of section~\ref{sec:timeflight}. We assume $\xi$ continuously differentiable $\forall t > 0$, there exists $T$ such that $\xi(t) \neq 0$, $\forall t> T$ and
\begin{equation}\label{hypothesis}
\lim_{t \to \infty} \xi(t) = 0\, .
\end{equation}
Thus, $\lim_{t \to \infty} \dot{\xi}(t) = 0$. We will show that unless some specific conditions are met, we have 
\begin{equation}\label{thelimit}
  \lim_{t \to \infty} \left|\frac{\dot{\xi}(t)}{\xi(t)}\right| = 0\,.
\end{equation}
To prove this, we define $y = \nicefrac{1}{t}$ and $g(y) = \xi(\nicefrac{1}{y})$. Then, we have $\dot{\xi}(t) = -y^2 g'(y)$. The limit \eqref{thelimit} is written as
\begin{equation}\label{glimit}
  \lim_{y \to 0} \left| \frac{y^2 g'(y)}{g(y)} \right| \,.
\end{equation}
We can express the denominator as 
\begin{equation}\label{glinha}
 g(y) = y \frac{g(y)-0}{y-0} = y \left(g'(y) + \xi (y) \right) \,,
\end{equation}
where $\lim_{y \to 0} \xi(y) = 0$. Inserting Eq.~\eqref{glinha} into \eqref{glimit}, we obtain
\begin{equation}
  \lim_{y \to 0} \left| \frac{y g'(y)}{g'(y) + \xi(y)}\right| = \lim_{y \to 0} \left| \frac{y}{1 + \frac{\xi(y)}{g'(y)}} \right| \,,
\end{equation}
which vanishes as $y \to 0$ unless $\lim_{y \to 0} \frac{\xi(y)}{g'(y)} = -1$. This is the first result.
 
In the special case
\begin{equation}\label{specialcase}
  \lim_{y \to 0} \frac{\xi(y)}{g'(y)} = -1
\end{equation}
we cannot tell the value of the limit \eqref{thelimit} by the method above. We show here that one of three cases may happen:

\begin{enumerate}[i.]
  \item $\lim_{t \to \infty} \left|\frac{\dot{\xi}(t)}{\xi(t)} \right| = 0$;
  \item $\lim_{t \to \infty} \left| \frac{\dot{\xi}(t)}{\xi(t)} \right| = L \neq 0$;
  \item $\lim_{t \to \infty} \left| \frac{\dot{\xi}(t)}{\xi(t)} \right| = \infty$.
 \end{enumerate}

We have written $\frac{g(y)}{y} = g'(y) + \xi(y)$. Thus, $\xi(y) = \frac{g(y)}{y} - g'(y)$ and we have $\frac{\xi(y)}{g'(y)} = \frac{g(y)}{y g'(y)} -1$. Therefore, Eq.~\eqref{specialcase} is equivalent to
\begin{equation}
  \lim_{y \to 0} \frac{g(y)}{y g'(y)} = 0 \,.
\end{equation}
We can build functions with this asymptotic property by making
\begin{equation}
  g'(y) = \frac{g(y)}{y h(y)} \, ,
\end{equation}
where $h(y)$ is any differentiable function such that 
\begin{equation}
  \lim_{y \to 0} h(y) = 0\,.
\end{equation}
This construction gives us $g(y)$ of the form
\begin{equation}
 g(y) = C \exp \left( \int \frac{\ud y}{y h(y)} \right)\,,
\end{equation}
which implies
\begin{align}
  \xi(t) =&\, C \exp \left[ - \int^t \frac{\ud t'}{t' h(\nicefrac{1}{t'})} \right] \,,\label{epsilonlimite}\\ 
 \dot{\xi}(t) =&\, -\frac{C}{t h(\nicefrac{1}{t})}\exp \left[ - \int^t \frac{\ud t'}{t' h(\nicefrac{1}{t'})} \right] \, , \\
 \left|\frac{\dot{\xi}(t)}{\xi(t)} \right| =&\, \frac{1}{t h(\nicefrac{1}{t})} \, . 
\end{align}
Therefore, $\lim_{t \to \infty} \left|\frac{\dot{\xi}(t)}{\xi(t)} \right|$ is finite, unless asymptotically $h(\nicefrac{1}{t}) < \mathcal{O}(\nicefrac{1}{t})$. The first derivative $\dot{\xi}$ will only dominate $\xi$ in the case $h(\nicefrac{1}{t}) < \mathcal{O}(\nicefrac{1}{t})$, which implies, by \eqref{epsilonlimite}, that $\xi(t) < \mathcal{O}(e^{-t})$, which shows that those cases always fall in case~\ref{epsilonpequeno}.
\section{How $\xi(t)$ determines the causal structure of gMcVittie spacetimes} \label{appendix:causal}

In this Appendix, we show the steps that prove Theorem \ref{theorem-rpos}. We have to consider separately the case $\dot{r}_- \to 0^+$ and $\dot{r}_-(t) \to 0^-$. The proof is analogous for both cases, the only difference being in the sign of some functions. We then specialize to the case $\dot{r}_-(t) > 0$ for $t$ sufficiently large. According to \eqref{dotrminus}, this corresponds to $\dot{\xi}(t) \to 0^+$ as $t \to \infty$.

It is important to note that we need an additional assumption over $\xi(t)$ here, compared to Theorem~\ref{oteorema}. We assume that $\dot{\xi} (t)$ has a finite number of roots, such that it reaches zero either from above or from below and does not oscillate indefinitely between positive and negative values. More precisely:
\begin{equation}\label{finiteroots}
  \exists \, T> 0 \quad \text{such that} \quad \dot{\xi}(t) \neq 0 \,, \, \forall \, t > T \,,
\end{equation}
which implies that there exists $T$ such that either $\xi(t) > 0$ and $\dot{\xi}(t)< 0$ for all $t > T$ or $\xi(t) < 0$ and $\dot{\xi}(t)> 0$ for all $t > T$. 

%

%

 
%
%
%


\begin{proposition}\label{dontcross-pos}
If $\dot{r}_-(t) > 0$ for all $t > T$, and if $r(t)$ is an ingoing null geodesics, governed by Eq. \eqref{t of r}, that satisfy $r(t_0) > r_-(t_0)$ for some $t_0 > T$, then we have $r(t) > r_-(t)$ for all $t > t_0$.  

\begin{proof}
 We define $d(t) = r_-(t) - r(t)$ between the horizon and the geodesic $r(t)$. Thus, for a given curve $r(t)$, $d(t_0) > 0$. From the definition of $r_-$, $\dot{r}(t) \to 0$ as $r \to r_-$, thus, for all $t > t_0$ there exist $\epsilon > 0$ such that $\dot{d}(t) > 0$ at $d(t) =  \epsilon > 0$. Therefore,  $d(t) > 0$ for all $t > t_0$.
\end{proof}

\end{proposition}

Proposition~\ref{dontcross-pos} implies that, if $\dot{r}_-(t) > 0$ for large times, ingoing null curves that are near $r_-$ but below it remain so for indefinitely large times and reach $r_\infty$ from below $r_-$. This automatically excludes the causal structure depicted in Fig. \ref{figurakaloper}, as in that case there is no ingoing null curve that reaches the limit surface from below $r_-$. However, for the curves that come from above, near the horizon we have $d(t) < 0$ and $\dot{d}(t) > 0$, which means that it is possible for ingoing null geodesics coming from below to cross the $r_-$ apparent horizon. If those curves do cross the horizon at a finite $t$, then they will reach the limit surface $r_\infty$ from below $r_-$, since they cannot cross $r_-$ back from above. Otherwise, if they do not cross $r_-$ at finite $t$, they obviously reach the limit surface from above. From this reasoning, we can predict that, when $\dot{r}_- > 0$ for large $t$, only the cases depicted in Figs.~\ref{figuragota} and \ref{figuralake} are possible. In other words, we need only focus our attention on geodesics approaching $r_-$ from above. The distinction between these two remaining scenarios may be stated as the following: if there exists $T$ such that, for any $t > T$, the ingoing geodesics just above $r_-$ fail to cross the horizon, then our causal structure is like that in Fig.~\ref{figuralake}. If there is no such $T$, in which case all ingoing geodesics coming from above the horizon do cross it in finite time, then our causal structure is that of Fig.~\ref{figurakaloper}.

The next proposition is very useful to simplify the analysis necessary to distinguish the remaining possible cases.



\begin{proposition}\label{rlimit}

Let $r(t)$ be an ingoing geodesic and $\dot{r}_-(t) > 0$.  If $r$ satisfies $r(t) > r_-(t)$ for all $t>t_0$, then
\begin{gather}
r (t) > r_{\infty}, \ \forall\, t > t_0 \,, \label{parte1}
\intertext{and}
\lim_{t \to \infty} r(t) = r_{\infty} \,. \label{parte2}
\end{gather}

\begin{proof}
 We refer the reader to the proof of Proposition III-1 of Ref.~\cite{daSilva:2012nh}, but exchanging incresing sequences by decreasing ones and vice-versa.
\end{proof}

\end{proposition}

Thanks to Proposition \ref{rlimit}, one only has to analyze geodesics crossing $r_\infty$, a fixed surface, eliminating the complication of having to consider the time-varying inner horizon $r_-(t)$. Moreover, it implies that the $r_\infty$ surface is an accumulating point for geodesics coming from above, as we state in the following Corollary.

\begin{corollary}\label{r-epsilon}

If $r (t)$ is an ingoing geodesic and $r(t_0) > r_-(t_0)$, then for all $\epsilon>0$ such that the events $(t, r_\infty + \epsilon)$ lies in the regular region $\forall t > t_0$, there exists $\bar{t} > t_0$ such that $r(\bar{t}) = r_{\infty} +\epsilon$.

\end{corollary}


This means that, given enough time, all geodesics either cross $r_\infty$ or reach values arbitrarily close to it. By Proposition \ref{rlimit}, every ingoing geodesic that never crosses $r_-(t)$ never reaches $r_\infty$. Conversely, if an ingoing geodesic does traverse $r_\infty$ in a finite time interval, then it eventually crosses $r_-$. Then, by studying only the neighborhood of $r_\infty$, we may tell if geodesics do or do not cross the apparent horizon. 
With all this in mind, we can use the approximations made in Sec.~\ref{sec:timeflight}, which are valid for large $t$ and $r$ near $r_\infty$, in order to use Eq.~\eqref{linearapprox} and again defining $z = r - r_\infty$, we obtain Eq. \eqref{linearapproxZ}.
%
Here we are interested in the error term of the approximation. We will use it to build approximations for the ingoing null curves we are studying. We call them $z^{\sigma}_{\pm}(t)$, which are curves that approximate $z(t)$ from above and below respectively, valid when close to $r_\infty$, that is, $z \leq \delta$. For an initial condition $z(t_0) = z_0 > 0$  and $0< z_0< \delta$, that is, above and close to the horizon, we are looking for two behaviors of the approximations $z^{\sigma}_{\pm}(t)$ for large $t$:
\begin{enumerate}
 \item $z^{\sigma}_-(t) > 0$ for all $t > t_0$. This guarantees that the ingoing geodesic does not cross the horizon in finite time.
 \item $z^{\sigma}_+(t) < 0$ for some $t > t_0$. Then, its guaranteed that the ingoing geodesic does cross the horizon at a finite time $ \bar{t} < t$.
\end{enumerate}


Now, using the fact that there exists $\sigma> 0$ such that
\begin{gather}
 \sigma z > \textit{o}(\delta) > -\sigma z  \, , \, \text{for} \quad z_0< z< 0 \, ,
\end{gather}
we build the approximations $z^{\sigma}_{\pm}(t)$ as solutions of
\begin{gather}
 \frac{\ud z^{\sigma}_{\pm}}{\ud t} = - \alpha H_0z^{\sigma}_{\pm} + R_\infty \xi(t) \pm \sigma z^{\sigma}_{\pm}\,, \nonumber\\ 
 z^{\sigma}_{\pm}(t_0) = z_0 \,,
\end{gather}
such that we obtain, by Gronwall's Lemma \cite{viterbo},
\begin{gather}
 z^{\sigma}_+(t) \geq z(t) \geq z^{\sigma}_-(t) \, , \qquad  t> t_0 \, ,
\end{gather}
\noindent
meaning that $z^{\sigma}_+ (t)$ approximate $z(t)$ from above and $z^{\sigma}_-(t)$ approximates $z(t)$ from below as we intended.

The solutions for $z^{\sigma}_\pm (t)$ are formally given by 
\begin{gather}
 z^{\sigma}_{\pm}(t) = e^{-\left(\alpha H_0\mp \sigma\right) t}\left\{R_\infty  \int_{t_0}^t e^{\left(\alpha H_0\mp \sigma\right) u} \xi(u) \ud u + Z_0 \right\} \, . \label{Zsolution+}
\end{gather}

We remark that $Z_0$ is positive and the integral term is negative, since $\xi(t) < 0$ for large $t$. Then we have to study the convergence of the functions $F^{\sigma}_\pm (t_0 , t)$  defined as
\begin{equation}
 F^{\sigma}_\pm (t_0 , t)= \int_{t_0}^t e^{\left(\alpha H_0\mp \sigma\right) u} \xi(u) \ud u < 0 \, ,
\end{equation}
\noindent

\begin{enumerate}

\item There exist $L_+(t_0) > 0$ and $\sigma> 0$ such that $\lim_{t \to \infty} F^{\sigma}_+(t_0,t) = L_+(t_0)>0$; \label{bounded}

\item There exist $\sigma>0$, such that $\lim_{t \to \infty} F_-(t_0,t) = + \infty$; \label{unbounded}

\end{enumerate}
\noindent
which correspond to the following scenarios:

Then, we conclude that

\begin{enumerate}
 \item If there exist $\sigma > 0$ such that $F^{\sigma}_+ (t_0, t) \to -\infty$ as $t \to \infty$, then there exist an approximation from above $z^{\sigma}_+ (t)$ such that the negative term of $z^{\sigma}_+(t)$ always can surpass the positive term and then $z(t)$ changes sign. This implies that all curves $z(t)$ changes sign and approach the limit surface from below and we obtain the structure depicted in fig.~\ref{figuragota}.
 \item If there exist $\sigma > 0$ such that $F^{\sigma}_-(t_0,t)$ converges to a value $L_- (t_0)< 0$ as $t \to \infty$, then, by a similar reasoning, there are approximations from below, $z_-(t)$ such that the positive term $Z_0$ is always larger in magnitude than the negative term, since we can make the integral term as small as we want by taking a larger $t_0$. Thus, in this case, we can say that there exists $T>0$, such that for $t_0 > T$, $z(t) > 0$ for all $t > t_0$. This correspond to the case where some curves approach the limit surface from below and some (the later ones) approach the limit surface from above. This correspond to the structure depicted in fig.~\ref{figuralake}.
\end{enumerate}

This proves half of the Theorem~\ref{theorem-rpos}. The proof of the case $\dot{r}_- \to 0^-$ is identical but for the orientation of the $z$-axis. More details in this kind of argument can be found in Ref. \cite{daSilva:2012nh}.

\bibliography{shortnames,referencias}

\end{document}